\documentclass[reprint, amsmath,amssymb,aps,twocolumn, prb,superscriptaddress]{revtex4-1}
\usepackage{amsfonts,amssymb}
\usepackage[subnum]{cases}
\usepackage{mathrsfs}
\usepackage{amsmath}
\usepackage[none]{hyphenat}
\usepackage{graphicx}
\usepackage{hyperref}    
\usepackage{bm}          
\usepackage{natbib}
\usepackage{soul} 
\usepackage{color}
\usepackage{longtable}
\usepackage{ textcomp }
\usepackage{verbatim}

\begin{document}

\title{Anomalous conductivity, Hall factor, magnetoresistance, and thermopower of accumulation layer in $\text{SrTiO}_3$}

\author{Han Fu}
\email{fuxxx254@umn.edu}
\affiliation{Fine Theoretical Physics Institute, University of Minnesota, Minneapolis, MN 55455, USA}
\author{K. V. Reich}
\affiliation{Fine Theoretical Physics Institute, University of Minnesota, Minneapolis, MN 55455, USA}
\affiliation{Ioffe Institute, St Petersburg, 194021, Russia}
\author{B. I. Shklovskii}
\affiliation{Fine Theoretical Physics Institute, University of Minnesota, Minneapolis, MN 55455, USA}
\date{\today}

\begin{abstract}
We study the low temperature conductivity of the electron accumulation layer induced by the very strong electric field at the surface of $\text{SrTiO}_3$ sample.
Due to the strongly nonlinear lattice dielectric response, the three-dimensional density of electrons $n(x)$ in such a layer decays with the distance from the surface $x$ very slowly as $n(x) \propto 1/x^{12/7}$. We show that when the mobility is limited by the surface scattering the contribution of such a tail to the conductivity diverges at large $x$ because of growing time electrons need to reach the surface. We explore truncation of this divergence by the finite sample width, by the bulk scattering rate, by the back gate voltage, or by the crossover to the bulk linear dielectric response with the dielectric constant $\kappa$. As a result we arrive at the anomalously large mobility, which depends not only on the rate of the surface scattering, but also on the physics of truncation. Similar anomalous behavior is found for the Hall factor, the magnetoresistance, and the thermopower.
\end{abstract}
\maketitle

\section{Introduction}

There is growing interest in the investigation of $\mathrm{ABO_3}$ perovskite crystals, which are important for numerous technological applications and show intriguing magnetic, superconducting, and multiferroic  properties \cite{Oxides_rev}.  Special attention \cite{Stemmer_STO,Zubko_oxides} is paid to heterostructures involving $\mathrm{SrTiO_3}$ (STO) which is a  semiconductor  with a band gap $E_g\simeq \mathrm{3.2~eV}$  \cite{Optical_absorbtion_STO} and a large dielectric constant $\kappa =2 \cdot 10^4$ at liquid helium temperatures. STO can be used as a building block for different types of devices, with reasonably large mobility \cite{Ohtomo_2004,Hwang_mobility}.

\begin{figure}
\includegraphics[width=0.8\linewidth]{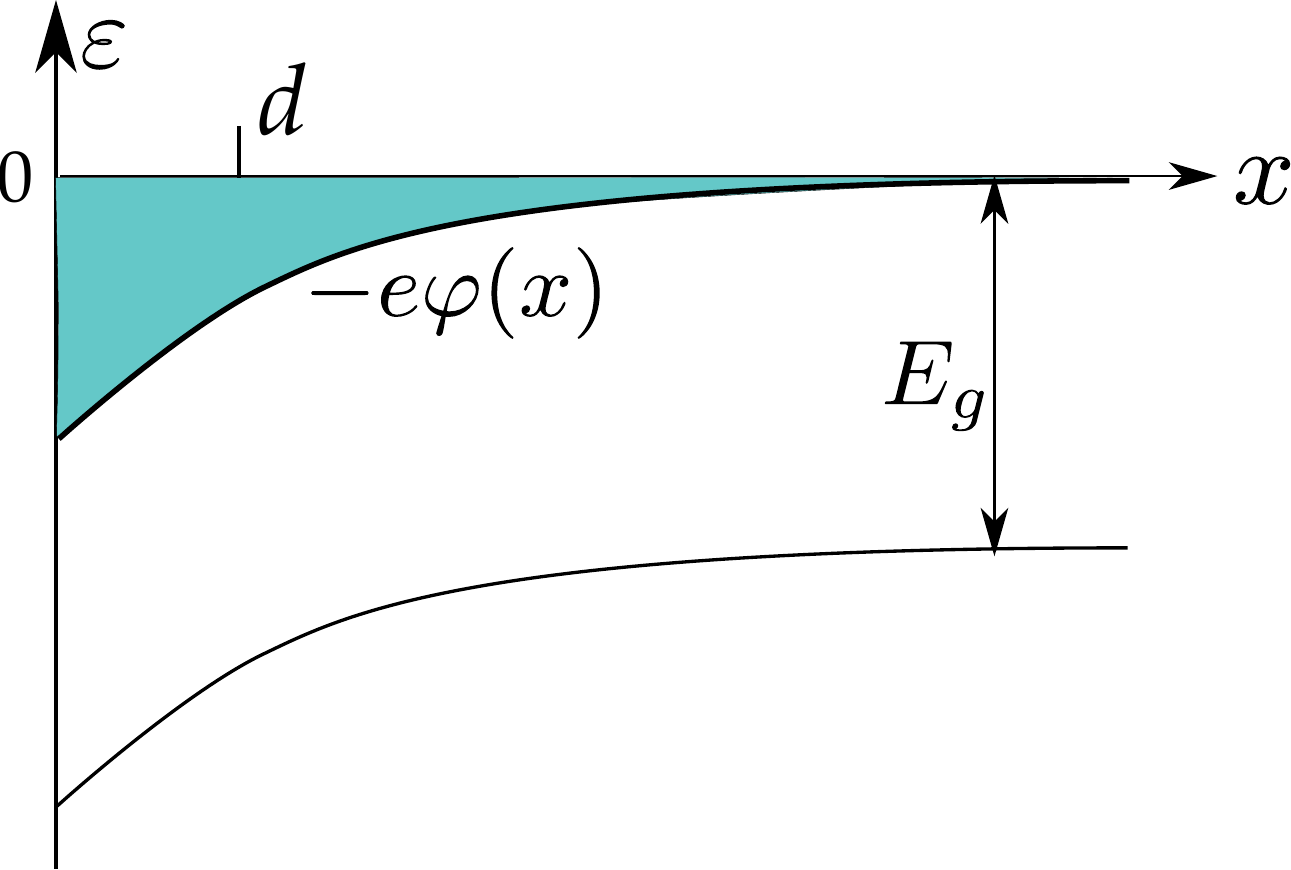}
\caption{(Color  online) Schematic electron potential energy $-e\varphi(x)$ diagram of an accumulation layer in a moderately $n$-doped STO where $x$ is the distance from the interface. Electrons (blue region) are attracted by an external induction $D_0$ applied at $x=0$. The characteristic width of the electron gas is $d$. In the bulk of STO the Fermi level   $\varepsilon_F $ is near the bottom of the conduction band.}
\label{fig:accumulation}
\end{figure}

Many devices are based on the accumulation layer of electrons near a heterojunction interface in a moderately $n$-type doped STO. For example, one can get an accumulation layer with two-dimensional (2D) concentration $N=3\times 10^{14}$ cm$^{-2}$ of electrons on the STO side of the GTO/STO heterojunction induced by the electric field resulting from the ``polar catastrophe" in GdTiO$_3$ (GTO) \cite{Stemmer_GdTO} (see Fig. \ref{fig:accumulation}). The role of GTO can also be played by perovskites LaAlO$_3$ \cite{Ohtomo_2004,Hwang_mobility,Stemmer_STO}, NdAlO$_3$, LaVO$_3$ \cite{LaVO_STO}, SmTiO$_3$, PrAlO$_3$, NdGaO$_3$ \cite{different_polar_STO}, LaGaO$_3$ \cite{LaGaO_STO}, and LaTiO$_3$ \cite{LaTO_STO}. One can  accumulate an electron gas using a field effect \cite{10_percent,Hwang_gate,Stemmer_concentration_interface}. In Refs. \onlinecite{induced_superconductivity,Gallagher_2014} the authors accumulated up to $10^{14} ~\mathrm{cm}^{-2}$ electrons on the surface of STO using ionic liquid gating. Inside bulk STO $\delta$-doping by large concentrations of donors can be used  to introduce two accumulation layers of electrons  \cite{delta_doped_stemmer,delta_doped_STO_Hwang,delta_doped_STO_Stemmer}.
Not surprisingly, the potential and electron density depth profiles in such devices have attracted a lot of attention \cite{Hwang_Xray,Hwang_PL,LAO_STO_Berreman,Stemmer_GdTO,MacDonald_theory,induced_superconductivity, abinitio_STO,abinitio_STO_2,distribution_LAO_STO,superconductivity_LAO_STO}.

In Ref. \onlinecite{RS}, authors calculated the three-dimensional (3D) electron density profile $n(x)$ of the accumulation layer with a large 2D density $N=\int_0^\infty n(x)dx$. To account for the nonlinear dielectric response in STO they used the Landau-Ginzburg free energy expansion \cite{Ginzburg_ferroelectrics, Landau_stat} while they described the degenerate electron gas with the Thomas-Fermi approximation \cite{Thomas_Fermi}. They arrived at the self-consistent potential $\varphi(x)$
\begin{equation}\label{eq:potential_nonlinear}
\varphi(x)=C_1\frac{e}{a}\left(\frac{a}{x+d}\right)^{8/7}
\end{equation}
and the electron concentration
\begin{equation}
n(x)=C_2\frac{1}{a^3}\left(\frac{a}{x+d}\right)^{12/7}\label{eq:nonlinear_concentration},
\end{equation}
where $a$ is the lattice constant, $d$ is the characteristic decay length of the electron density
\begin{equation}
d=C_3a\left(Na^2\right)^{-7/5}\label{eq:decay_length}.
\end{equation}
Here $C_1,\,C_2,\,C_3$ are dimensionless constants of order unity which can be found in Ref. \onlinecite{RS}. Note that $n(x)$ has an unusually long tail with a weak 12/7 power law dependence. This form of $n(x)$ seems to be in agreement with experimental data \cite{RS}.

In this paper we assume that the scattering of electrons is on the surface roughness or on some ions near the interface. Therefore, the scattering rate of electrons in the body of the distribution $n(x)$ is much larger than that of electrons in the tail of $n(x)$ due to the large travel time to the surface of the tail electrons. As a result the tail contribution to different kinetic coefficients diverges. In particular, this leads to the anomalously large  mobility, Hall factor, magnetoresistance, and thermopower, which depend on the truncation mechanism of the divergences. Similar anomalies were predicted for silicon MOSFETs at high temperatures~\cite{Entin}.

The interplay between contributions from tail and body electrons to kinetic coefficients can be interpreted as the existence of two types of carriers. This option has been widely discussed recently for the data on the linear and nonlinear Hall effect \cite{Zeitler}, on the inconsistency between electron concentrations measured by the Hall effect and the Shubnikov-de Haas effect \cite{Ariando}, and on the difference between ac and dc transport results \cite{Lee_Hwang}.

The paper is organized as follows. In Sec. \ref{sec:conductivity}, we demonstrate the divergence of the conductivity due to the long tail of $n(x)$ and study its cutoff by several truncating mechanisms. In Sec. \ref{sec:other}, we do similar analysis for the Hall factor. In Sec. \ref{sec:thermo}, we study the anomalous magnetoresistance and thermopower. In Sec. \ref{sec:dis}, we discuss applicability of our results. We conclude in Sec. \ref{sec:con}.

\section{Conductivity}\label{sec:conductivity}
In Introduction, we described the electron distribution in an accumulation layer induced in STO-based heterointerfaces which has a long tail $n(x)\propto x^{-12/7}$.
If we ignore the scattering of electrons by bulk impurities, the low temperature mobility of the accumulation layer in STO is limited by the surface (interface) scattering. Since the time an electron originally at the Fermi level of distance $x$ spends on the journey to the surface is $\sim x/v_F(x)$ where $v_F(x)\sim \hbar n(x)^{1/3}/m^*\propto x^{-4/7}$ is the Fermi velocity and $m^*$ is the effective electron mass, we get the corresponding relaxation time
\begin{equation}\label{eq:tau_surface_nonlinear}
\tau(x)=\tau_s\left(\frac{x}{d}\right)^{11/7},
\end{equation}
where $\tau_s\equiv\tau(d)$ is the surface scattering related relaxation time of electrons in the body of distribution \eqref{eq:nonlinear_concentration}. The spatially varying relaxation time $\tau(x)$ has to be averaged to calculate the surface conductivity. Usually, in bulk semiconductors when there are different kinds of carriers, e.g., electrons from two bands with the same effective mass $m^*$, the 3D conductivity is $e^2n\bar\tau/m^*$, where $n=n_1+n_2$ is the total 3D concentration of different carriers, and the averaged relaxation time is \cite{Blatt}
\begin{equation}\label{conductivty_1}
\bar\tau=\frac{n_1\tau_1+n_2\tau_2}{n_1+n_2}.
\end{equation}
Here the subscripts refer to the concentrations and relaxation times of the two different carriers. One can generalize Eq. \eqref{conductivty_1} to our case where electrons at different $x$ have different relaxation times and thus behave as if they are different carriers. The total 2D conductivity is then
\begin{equation}\label{eq:conductivity_2}
\begin{aligned}
\sigma=\frac{e^2N\left<\tau\right>}{m^*}
\end{aligned}
\end{equation}
where similarly to Eq. \eqref{conductivty_1}, we have here
\begin{equation}\label{eq:average_tau}
\left<\tau\right>=\frac{\int_0^L dx\, n(x)\tau(x)}{\int_0^L dx\,n(x)}=\frac{\int_0^L dx\, n(x)\tau(x)}{N}.
\end{equation}
Here $N$ is the total 2D concentration of electrons. Below we always understand the averaging $\left<\dots\right>$ in the way of Eq. \eqref{eq:average_tau}.
Using Eqs. \eqref{eq:nonlinear_concentration} and \eqref{eq:tau_surface_nonlinear}, we then obtain
\begin{equation}\label{eq:conductivity_result}
\sigma=\sigma_s\left(\frac{L}{d}\right)^{6/7}
\end{equation}
where $\sigma_s=Ne\mu_s$, $\mu_s=e\tau_s/m^*$ is the electron mobility in the body of $n(x)$ distribution at $x\le d$. We see that both $\left<\tau\right>$ and $\sigma$ diverge in the limit $L\rightarrow \infty$. This is why we had to introduce a finite truncation length $L$ to the electron density tail. It can be specified for several possible truncation mechanisms: i) the finite width of the STO sample, ii) a finite bulk scattering rate, and iii) the nonlinear-linear dielectric response transition. The smallest of these values is to be substituted into Eq. \eqref{eq:conductivity_result}.

\emph{Finite sample width $W$} --- For a sample with a relatively small width $W$, for example, GTO/STO/GTO structures with the STO layer of width $W$, the resulting conductivity is
\begin{equation}\label{eq:finitewidthconductivity}
\sigma=\sigma_s\left(\frac{W}{d}\right)^{6/7}
\end{equation}
with $L$ in Eq. \eqref{eq:conductivity_result} substituted by $W\gg d$. The expression of the relaxation time $\tau_s$ depends on the surface scattering mechanism. 

\emph{Bulk scattering} --- Let us now consider the large $W$ case and assume that the bulk relaxation time $\tau_b$ does not depend
on the electron concentration \footnote{A mobility independent of the electron concentration of course means that in the tail far enough from the surface where $n(x)$ becomes very small electrons should get localized. Indeed, it is known that in the bulk STO samples the localization happens at concentration $n_c \sim 3\times10^{16}$ cm$^{-3}$ according to Ref. \onlinecite{Chris_STO_doped}. In this paper we
deal with accumulation layers with much larger near-the-interface concentrations $n(0) \sim 10^{20} $cm$^{-3}$ so that
other mechanisms are assumed to truncate the conductivity or the Hall factor before $n(x)$ reaches $n_c$.}. (We justify this assumption in Sec. \ref{sec:dis}.)
Assuming that $\tau_b \gg \tau_s$ we can find such a distance $x=L_1$ that the relaxation time Eq. \eqref{eq:tau_surface_nonlinear} due to surface scattering and $\tau_b$ are equal
\begin{equation}
\tau_s\left(\frac{L_1}{d}\right)^{11/7}=\tau_b.
\end{equation}
This gives
\begin{equation}\label{eq:truncate_bulk}
L_1=d\left[\frac{\tau_b}{\tau_s}\right]^{7/11}\gg d.
\end{equation}
At $x\gg L_1$, the total relaxation time $\left[\tau^{-1}(x)+\tau^{-1}_b\right]^{-1}\approx \tau_b$ is constant and the conductivity converges. Thus, substituting $L_1$ for $L$ in Eq. \eqref{eq:conductivity_result}, we get
\begin{equation}\label{eq:2D_conductivity_bulk_truncate}
\sigma=Ne\mu_s\left[\frac{\tau_b}{\tau_s}\right]^{6/11}=Ne\mu_b^{6/11}\mu_s^{5/11}
\end{equation}
where $\mu_b=e\tau_b/m^*\gg\mu(d)$ is the electron mobility due to bulk impurity scattering. This gives the final value of $\sigma$ only for relatively large samples when the width $W>L_1$ and $\sigma$ obtained from Eq. \eqref{eq:2D_conductivity_bulk_truncate} is smaller than Eq. \eqref{eq:finitewidthconductivity}. A remarkable feature of Eq. \eqref{eq:2D_conductivity_bulk_truncate} is that the final mobility depends on both the surface and the bulk scattering and is close to the geometrical average $\left[\mu_b \mu_s\right]^{1/2}$.

\emph{Crossover to linear dielectric response} --- The electric field of the accumulation layer decays with $x$ as $1/x^{15/7}$ and eventually becomes so small that the dielectric response of STO becomes linear with the large dielectric constant $\kappa$. According to Ref. \onlinecite{RS}, this happens when $x$ reaches \begin{equation}\label{eq:truncate_linear}
L_2=C_4a\kappa^{7/10}\gg d,
\end{equation}
where $C_4$ is of order unity \cite{RS}. At $x\gg L_2$, the 3D electron concentration is
\begin{equation}\label{eq:concentration_linear}
n(x)\simeq C_5 \frac{a_B^3}{x^6}
\end{equation}
where $C_5\approx 442$ and the conductivity converges for this density profile. This means that at $L_2\ll W,\,L_1$, we can get the conductivity substituting $L_2$ for $L$ in Eq. \eqref{eq:conductivity_result}. As a result,
\begin{equation}\label{eq:conductivity_L2}
\sigma=\sigma_s\left(\frac{L_2}{d}\right)^{6/7}\simeq \sigma_s\left(Na^2\right)^{6/5}\kappa^{3/5}.
\end{equation}
In this case, of course, in its range of validity Eq. \eqref{eq:conductivity_L2} gives a smaller $\sigma $ than both Eqs. \eqref{eq:finitewidthconductivity}
and \eqref{eq:2D_conductivity_bulk_truncate}.

\section{Hall factor}\label{sec:other}
In this section we discuss effects of a weak magnetic field $B$ on the conductivity tensor: the Hall effect.
It is known that the 3D Hall constant is $r_H/nec$, where the Hall factor according to Ref. \onlinecite{Blatt} is
\begin{equation}\label{eq:Hall_constant}
r_H=\frac{(n_1\tau_1^2+n_2\tau_2^2)(n_1+n_2)}{(n_1\tau_1+n_2\tau_2)^2}
\end{equation}
for two kinds of carriers with the same effective mass but different relaxation times labeled by subscripts $1,\,2$. So again, we can generalize this result to our case where electrons at different positions play the role of carriers with different $\tau$. The Hall factor is then
\begin{equation}\label{eq:Hall_factor}
r_H=\frac{\left<\tau^2\right>}{\left<\tau\right>^2}
\end{equation}
where the averaging is weighed by the electron 2D concentration ratio $dx\, n(x) /N$ following the form of Eq. \eqref{eq:average_tau}. Using Eq. \eqref{eq:tau_surface_nonlinear} one can see that when $\left<\tau \right>$ diverges $\left<\tau^2 \right>$ diverges even stronger.
Therefore below we deal with the truncation of both divergences.

For relatively thin STO samples where $W\ll L_1,\,L_2$ and $L_1,\,L_2$ are given respectively by Eqs. \eqref{eq:truncate_bulk} and \eqref{eq:truncate_linear}, both divergences of $\left<\tau^2\right>$ and $\left<\tau\right>$ are cut by $W$. According to Eq. \eqref{eq:tau_surface_nonlinear}, we get
\begin{equation}\label{eq:Hall_surface}
\left<\tau^2\right>=\tau_s^2\left(\frac{W}{d}\right)^{17/7},\,\left<\tau\right>=\tau_s\left(\frac{W}{d}\right)^{6/7}
\end{equation}
so the Hall factor is
\begin{equation}\label{eq:Hall_factor_1}
r_H=\left(\frac{W}{d}\right)^{5/7}.
\end{equation}
When the STO sample width is larger, i.e., $W\gg L_1$, the bulk scattering becomes important before the electron density vanishes. From Sec. \ref{sec:conductivity}, we know that $\left<\tau\right>$ stops diverging at this point. Meanwhile, due to the constant relaxation time $\tau_b$ at $x>L_1$, $\left<\tau^2\right>$ also stops diverging, so we arrive at
\begin{equation}\label{eq:Hall_factor_2}
r_H=\left(\frac{L_1}{d}\right)^{5/7}
\end{equation}
with $L_1$ here playing the role of $W$ in Eq. \eqref{eq:Hall_factor_1}.
This result is valid only when the dielectric response is nonlinear at all $x<L_1$, i.e., $L_2\gg L_1$. When $L_2\ll L_1,\,W$, the divergence of $\left<\tau\right>$ stops at $x=L_2$ but $\left<\tau^2\right>$ continues diverging even after this point where $n(x)$ crosses over to $\propto 1/x^6$. Indeed, in this case instead of Eq. \eqref{eq:tau_surface_nonlinear} we get \begin{equation}\label{eq:tau_surface_linear}
\tau(x)=\frac{x}{v_F(x)}\propto x^3
\end{equation}
where $v_F(x)\propto n(x)^{1/3}\propto 1/x^2$. As a result $\left<\tau^2\right>\simeq \tau_s^2(L_2/d)^{17/7}(L/L_2)$. To truncate this new divergence we should use the finite sample width $W$ or the bulk scattering to obtain $L$. However, one should note that the position where the bulk scattering dominates changes from $L_1$ to $L_1'$ now due to the new dependence of $\tau(x)$ Eq. \eqref{eq:tau_surface_linear}. Since at $x>L_2$, $\tau(x)=\tau(L_2)\left(x/L_2\right)^3$ where $\tau(L_2)=\tau_s\left(L_2/d\right)^{11/7}$ given by Eq. \eqref{eq:tau_surface_nonlinear} at $x=L_2$, we now get $\tau(x)=\tau_b$ at $x=L_1'$ and
\begin{equation}\label{eq:truncate_bulk_new}
\begin{aligned}
L_1'=&d\left(\frac{\tau_b}{\tau_s}\right)^{1/3}\left(\frac{L_2}{d}\right)^{10/21}.\\
\end{aligned}
\end{equation}
At $L_2\ll L_1'\ll W$, we have
\begin{equation}\label{eq:Hall_bulk_1}
\begin{aligned}
\left<\tau^2\right>=&\,\tau_s^2\left(\frac{L_2}{d}\right)^{17/7}\left(\frac{L_1'}{L_2}\right),\\
\left<\tau\right>=&\,\tau_s\left(\frac{L_2}{d}\right)^{6/7}
\end{aligned}
\end{equation}
and
\begin{equation}\label{eq:Hall_factor_3}
r_H=\left(\frac{L_2}{d}\right)^{5/7}\left(\frac{L_1'}{L_2}\right).
\end{equation}
At $L_2\ll W\ll L_1'$, we get
\begin{equation}\label{eq:Hall_factor_4}
r_H=\left(\frac{L_2}{d}\right)^{5/7}\left(\frac{W}{L_2}\right)
\end{equation}
with $W$ substituting for $L_1$ in Eq. \eqref{eq:Hall_factor_3}. Obviously, Eqs. \eqref{eq:Hall_factor_1}, \eqref{eq:Hall_factor_2}, \eqref{eq:Hall_factor_3}, and \eqref{eq:Hall_factor_4} are valid only for $r_H \gg 1$.

One should note that above results are valid only for the weak enough magnetic field. This means that for all relevant values of $x$, the inequality $\omega _c\tau(x)\ll 1$ is fulfilled, where $\omega_c=eB/m^*c$ is the cyclotron frequency. Let us now consider the large $B$ case when $\omega_c\tau(x)=1$ already at $x=L_3< W,\,L_1\, (L_1'),\,L_2$. Using Eq. \eqref{eq:tau_surface_nonlinear}, we obtain
\begin{equation}
\omega_c\tau_s\left(\frac{L_3}{d}\right)^{11/7}=1
\end{equation}
and
\begin{equation}\label{eq:truncate_B}
L_3=\frac{d}{\left(\omega_c\tau_s\right)^{7/11}}\propto\frac{1}{B^{7/11}}.
\end{equation}
This is the truncation length of the electron density tail by the magnetic field. The resulting Hall factor for $L_3\gg d$ is
\begin{equation}\label{eq:Hall_factor_5}
r_H=\left(\frac{L_3}{d}\right)^{5/7}\propto \frac{1}{B^{5/11}}.
\end{equation}
For simplicity we skip analysis of intermediate magnetic fields where some of other truncation lengths are smaller than $L_3$.

\section{Magnetoresistance and thermopower}\label{sec:thermo}

\emph{Magnetoresistance} ---
When a weak magnetic field $B$ is applied normal to the interface, the resistivity $\rho$ of the accumulation layer changes by $\Delta \rho=\rho(B)-\rho(0)$ where $\rho(B)$ is the magnetoresistance. According to Ref. \onlinecite{Blatt} the magnetoresistance ratio at small $B$ is
\begin{equation}\label{eq:magneto}
\frac{\Delta\rho}{\rho}=\frac{\left(\left<\tau^3\right>\left<\tau\right>-\left<\tau^2\right>^2\right)\omega_c^2}{\left<\tau\right>^2}.
\end{equation}
We can repeat previous analysis for the magnetoresistance and summarize the final results in Table. \ref{tab:2}.
\begin{table}[h]
\caption{\label{tab:2} Magnetoresistance ratio $\Delta\rho/\rho$ in units of $\tau_s^2\omega_c^2$ and thermopower $S$ in units of $k_B^2T/e^2\varphi(d)$  at different truncation situations where $\tau_s$ is the electron relaxation time due to surface scattering in the body of $n(x)$, $\omega_c=eB/m^*c$ is the cyclotron frequency, $\varphi(d)\simeq \left(Na^2\right)^{8/5}e/a$ is the electric potential in the body of $n(x)$ according to Eq. \eqref{eq:potential_nonlinear}. Here $W$ is the width of the STO sample, $d,\,L_1,\,L_2,\,L_1'$ are given by Eqs. \eqref{eq:decay_length}, \eqref{eq:truncate_bulk}, \eqref{eq:truncate_linear}, and \eqref{eq:truncate_bulk_new}, respectively.}
\begin{ruledtabular}
\renewcommand{\arraystretch}{1.5}
\begin{tabular}{ c| c |c}
& $\Delta\rho/\rho$ &$S$ \\ \hline
$W\ll L_1,\,L_2$ &$(W/d)^{22/7}$ & $(W/d)^{8/7}$\\ \hline
$L_1\ll L_2 \ll W$& $(L_1/d)^{22/7}$&$(L_1/d)^{8/7}(L_2/L_1)^{3/7}$\\ \hline
$L_1\ll W\ll L_2$& $(L_1/d)^{22/7}$&$(L_1/d)^{8/7}(W/L_1)^{3/7}$\\ \hline
$L_2\ll L_1'\ll W$ &$(L_2/d)^{22/7}(L_1'/L_2)^4$  &$(L_2/d)^{8/7}(L_1'/L_2)^2$\\ \hline
$L_2\ll  W\ll L_1'$&$(L_2/d)^{22/7}(W/L_2)^4$&$(L_2/d)^{8/7}(W/L_2)^2$
\end{tabular}
\end{ruledtabular}
\end{table}

\emph{Thermopower} --- Another important property of the system is its thermopower $S$ which is the ratio of the induced electric field to the temperature gradient \cite{Blatt}
\begin{equation}\label{eq:thermopower_final}
\begin{aligned}
S
\simeq&\frac{k_B^2T}{e}\frac{\left<\tau/\varepsilon\right>}{\left<\tau\right>}\propto\frac{\left<\tau/n^{2/3}\right>}{\left<\tau\right>}
\end{aligned}
\end{equation}
where $k_B$ is the Boltzmann constant, $T$ is the temperature.
Here $\left<\tau/n^{2/3}\right>$ is strongly divergent and only at $x>L_1\, (L_1'),\,L_2$ or $x>W$ can it stops diverging. The results are shown in Table. \ref{tab:2}.

\section{Discussion}\label{sec:dis}
\emph{Relaxation time approximation} ---
Although all the scaling derivations of transport properties of the accumulation layer in STO in previous sections
were based on the relaxation time approximation, they can be justified by solving the Bolzmann kinetic equation.
As we demonstrated above, these transport properties are dominated by a small fraction of the tail electrons. In this case, the collision term in the Boltzmann equation is dominated by the relaxation
of this particular small fraction of electrons and therefore can be reduced to the relaxation time approximation.

\emph{Fermi level in the bulk of STO} ---
For simplicity we assumed that the bulk of STO is only lightly doped by donors so that the Fermi level in the bulk STO coincides with the conduction band bottom and the electron concentration tends to zero at large $x$ according to Eq. \eqref{eq:nonlinear_concentration}.  In this case, at $T=0$  the bulk of STO is insulating and does not contribute to the surface conductivity. Actually STO crystals as grown are believed to be strongly compensated~\cite{Chris_STO_doped} so that the Fermi level is in the STO band gap. This does not affect the accumulation layer structure because the conduction band bottom acquires its bulk position only at the distance comparable with the screening radius of thermally activated electrons which is exponentially large at low temperatures. When the bulk of STO has excessive acceptors with small concentration $n_A$ so that our accumulation layer becomes the inversion layer, even though the width of the hole depletion layer is very large, its total surface charge is much smaller than the electron surface charge $N$. In this case, acceptors do not affect the electron distribution $n(x)$ and all our results above are valid \footnote{More caution is required if the bulk of STO is heavily doped by donors and, therefore, has a finite concentration of degenerate electron gas. This is easily achievable because the Bohr radius of a donor $a_B = \hbar^2\kappa/m^*e^2 \sim 1000 $nm$ \gg a$. As a result the bulk is conducting and the surface conductivity of the accumulation layer should be defined as a difference between conductivity of the sample in strong applied electric field and without it. Also, the linear screening radius of the bulk electron gas truncates the accumulation layer, but because of the large dielectric constant this happens at a distance much larger than other truncation lengths. Thus, even in this case our theory remains valid.}.

\emph{Effect of back gate} --- If an STO sample with width $W$ has a back gate, one can apply to it a voltage $V$.
When $V< 0$ and $|V|$ is large enough, the back gate induced electric field $E= -V/W$ can squeeze the electron gas truncating the tail at a new distance $X_m(|V|)\ll W$. To find $X_m(|V|)$, we match electric fields at this point, i.e., $E(X_m) = - d\varphi/dx = -V/W$.
Using Eq. \eqref{eq:potential_nonlinear}, we arrive at $X_m \simeq a (|V|a^{2}/We)^{-7/15}$,
which is valid if $W \gg X_m(|V|) \gg d$.
Substituting this $X_m$ for $W$ into Eqs. \eqref{eq:finitewidthconductivity} and \eqref{eq:Hall_factor_1}, we arrive at
\begin{equation}
\sigma = \sigma_s \left(\frac{a}{d}\right)^{6/7} \left(\frac{|V|a^{2}}{We}\right)^{-2/5}
\end{equation}
and
\begin{equation}
r_H = \left(\frac{a}{d}\right)^{5/7} \left(\frac{|V|a^{2}}{We}\right)^{-1/3}.
\end{equation}

\emph{Applicability of Thomas-Fermi approach} --- All our results are based on Eq. \eqref{eq:nonlinear_concentration} for the electron density distribution, which was derived in the Thomas-Fermi (TF) approximation. Here we discuss the applicability of such an approximation. The TF approximation works if the potential varies at distances much larger than the electron wave length, or more exactly, when the TF parameter $k_F x \gg 1$. We showed \cite{RS} that $k_F d \simeq 3$ even at the very large $N=0.5~ a^{-2}$ achieved in the GTO/STO heterostructure. Generally speaking at $x \gg d$, where $k_F\sim n(x)^{1/3}\sim a^{-3/7} x^{-4/7}$ we get that the TF parameter $k_F x \sim (x/a)^{3/7}\gg 1$ and grows with $x$. It reaches its maximum value $\kappa^{3/10}$ at $x=L_2$, where the crossover to the linear dielectric response happens. At $x > L_2$ Eq. \eqref{eq:concentration_linear} gives $k_F\propto 1/x^2$ so that the TF parameter $k_Fx$ decreases as $1/x$ and at $x=a_B$ becomes of the order of 1. Here $a_B = \hbar^2\kappa/m^*e^2$ is the Bohr radius in STO. The accumulation layer terminates at $x = a_B$ so that $a_B$ is another truncation length, which we have not considered in Secs. \ref{sec:conductivity}, \ref{sec:other}, and \ref{sec:thermo} because in STO $a_B \sim 1000$ nm is larger than all other truncation lengths. Thus, the use of the TF approach is well justified.

\emph{Bulk relaxation time} --- In compensated STO at low temperatures electrons are scattered by charged donors and acceptors with total concentration $n_i \sim 5 \times 10^{18}$ cm$^{-3}$. We assumed above that the resulting bulk relaxation time $\tau_b$ does not depend on the electron concentration $n$. We can justify this assumption by appealing to experimental data summarized in Ref.~\onlinecite{Chris_STO_doped,Kamran}. It was shown there that in STO samples intentionally heavily doped by Nb donors at the level of $10^{18} < n < 4\times 10^{20}$ cm$^{-3} $ on the top of existing $n_i$ donor
and acceptors their 3D conductivity weakly depends on the 3D electron concentration of electrons $n$. This indicates that $\tau_b \propto n^{-1}$ when the scattering happens on donors of concentration $n$. This means that $\Sigma v_F$ does not depend on $n$. Here  $\Sigma$ is the scattering crosssection and $v_F$ is the Fermi velocity. Returning to undoped STO samples with $n_i$ donor and acceptors as scatterers we see that $\tau_b = (n_i \Sigma v_F)^{-1}$ does not depend on $n$.

\emph{Surface relaxation time} ---
Above, we have not specified the relaxation time $\tau_s$ of electrons in the body of the electron distribution \eqref{eq:nonlinear_concentration} due to the surface scattering and the corresponding mobility $\mu_s$. They can be limited by scattering on ionized donors and surface roughness. First, let us imagine that the surface has ionized impurities with the 2D concentration $N_s$. The effective bulk concentration of the scattering centers is $N_s/d$.  From the experimental data \cite{Chris_STO_doped,Kamran}, we know that $\Sigma v_F$ weakly depends on the electron concentration $n$, which leads to the mobility
$$
\mu_s\simeq\frac{e}{\hbar N_s} \frac{d}{a} \simeq\frac{e}{\hbar N_s} \frac{1}{(Na^2)^{7/5}}.
$$
There are many reasons for the existence of charged impurities near the surface. For example, it is believed, that the interface $\mathrm{LaAlO_3/SrTiO_3}$ has a large number of charged impurities with the 2D concentration $N_s$ and the electron surface concentration $N \neq N_s$  due to redistribution of ions near the interface \cite{Yu_new_mechanism}. Also, due to the discreteness of ions the gating of STO by ionic liquid is equivalent to the introduction of random Coulomb centers near the surface of STO as was shown in the case of Si \cite{Tardella,JJ_1,*JJ}.

At even larger concentration the mobility is limited by the surface roughness. We considered this case in details for the accumulation layer without nonlinear dielectric response \cite{roughness}. We assumed that the scattering occurs on the islands with typical diameter $D$ and height $\sim a$. We arrived at that the relaxation time is:
$$ \tau_s = \frac{m^*}{\hbar} \dfrac{dD}{k_F^2a^2}
  \begin{cases}
    D^{-3}k_F^{-3}   & \quad \text{if } k_F^{-1} \gg D \\
    1  & \quad \text{if } k_F^{-1} \ll D.\\
  \end{cases}
$$
For the nonlinear dielectric response case we get the same result in terms of $k_F\simeq \left[\,n(0)\,\right]^{1/3}$ and $d$. Now using the relationship between $d$, $k_F$, and $N$ for the nonlinear dielectric response Eqs. \eqref{eq:nonlinear_concentration} and \eqref{eq:decay_length} we arrive to the corresponding mobility:
$$ \mu_s = \frac{e}{\hbar} a^2
  \begin{cases}
    \dfrac{a^2}{D^2} (Na^2)^{-27/5}   & \quad \text{if } Na^2<(D/a)^{-5/4} \\
    \\
    \dfrac{D}{a} (Na^2)^{-3}  & \quad \text{if } Na^2>(D/a)^{-5/4}\\
  \end{cases}
$$

\emph{Beyond isotropic effective mass approximation} --- In this paper following Ref. \onlinecite{RS} we assumed that the electron spectrum at the bottom of the conduction band of STO can be approximated by the single isotropic band with the effective mass $m^*$. Actually, near the conduction band bottom of STO are three degenerate bands formed by $xy$, $xz$ and $yz$ Ti $d$-orbitals, which are anisotropic with the heavy mass direction along the $z,\,y,$ and $x$ axes, respectively. The splitting of these bands by the spin-orbit interaction \cite{Mazin_band_structure} can be ignored at relatively large electron concentrations $10^{19} < n < 10^{22}$ cm$^{-3}$ which we are interested in for accumulation layers with very large surface concentration $N$. Indeed, at electron concentrations larger than $10^{19}$ cm$^{-3}$, all energy bands are almost equally occupied so that the effective mass measured by the specific heat $m^* \simeq 1.8~ m$ \cite{Effective_mass_STO} does not change with $n$. One should note that $m^*$ describes the total density of states of all three bands. Our TF theory of the accumulation layer uses only the density of states. Thus, it is valid to use the effective mass $m^*$ when the TF criterion $k_F x \gg 1$ is fulfilled for all bands at all $x\geq d$. In this case, our Eq. \eqref{eq:nonlinear_concentration} is justified for both the body and the tail of the $n(x)$ distribution. When $N \ll 1/2a^2$ the electron distribution $n(x)$ is so wide ($d\gg a$) that the TF criterion is easily fulfilled for all bands. However, for the largest concentration $N = 1/2a^2$, where $d$ becomes comparable with the lattice constant the two bands with the light mass along the $x$ axis may only marginally satisfy the TF criterion. Near $x=0$, this depletes their contribution to the density of states and reduces the maximum value of $n(x)$. However, in the tail the TF criterion is still valid. Thus, the tail of $n(x)$ which plays the major role in this paper still follows Eq. \eqref{eq:nonlinear_concentration}. This conclusion agrees with numerical results for $n(x)$ obtained for $N = 1/2a^2$ in Refs. \onlinecite{RS,abinitio_STO_2,distribution_LAO_STO,Chen_Hanghui}.

\emph{Beyond STO} -- Above we dealt with accumulation layers in STO where the linear dielectric constant is very large and and dielectric response is strongly nonlinear. Our results are directly applicable to the very similar KTaO$_3$ and CaTiO$_3$ and to other materials with very large dielectric constant. The similar approach is also applicable to accumulation layers with large concentration of electrons in semiconductors with unremarkable dielectric properties such as Si \cite{Tardella, JJ_1,*JJ} or ZnO \cite{ZnO}. In such a crystal the dielectric response is linear and the electron concentration at low temperatures behaves as
\begin{equation}
n(x) = C_5 \frac{a_B^{3}}{(d_1 +x)^{6}}, \label{X}
\end{equation}
where $d_1 = a_B/(Na_{B}^2)^{1/5}$ is the new decay length of the electron distribution from the surface, $N$ is the total 2D electron density, $a_B$ is the Bohr radius of the semiconductor. In this case, using Eqs. \eqref{eq:tau_surface_linear} and Eq. \eqref{eq:concentration_linear} we arrive at the converging conductivity. However, the Hall factor, the magnetoresistance, and the thermopower diverge. For a weakly doped uncompensated bulk crystal with large width $W$ where the bulk relaxation time $\tau_b$ provides a large truncation length, the divergence is cut by the failure of the TF approximation at $x=a_B$ similarly to the termination of the standard TF atom electron density. The results for the Hall factor, the magnetoresistance, and the thermopower then are
\begin{equation}\label{eq:Hall_factor_linear}\begin{aligned}
r_H\, = &\left(Na_B^{2}\right)^{1/5},\\
\, \frac{\Delta\rho}{\rho} \,=&\left(\omega_c \tau_s\right)^{2} (Na_B^{2})^{4/5},\\
S\,\,=&\frac{k_B}{e}\left[\frac{k_BT}{e\varphi(d)}\right] \left(Na_B^{2}\right)^{2/5},\\
\end{aligned}
\end{equation}
where $\tau_s$ is the surface scattering relaxation time of electrons in the body of electron distribution, $e\varphi(d)=\left(e^2/\kappa_1a_B\right)\left(Na_B^2\right)^{4/5}$ is the depth of the Fermi sea near the surface \cite{RS} and $\kappa_1$ is the dielectric constant of the semiconductor. In Si and ZnO one can achieve $Na_B^{2}\sim 5$ so that one can see substantial anomalies.

\section{Conclusion}\label{sec:con}

In this paper, we have discussed transport properties of electron accumulation layers induced by a very strong electric field on the surface of the STO crystal. Due to the strongly nonlinear dielectric response of STO the electron density in an accumulation layer has a very compact body and a long slowly decaying tail. If in the body electrons are strongly scattered by the surface while in the tail electrons need a long time to reach the surface, the tail electrons run away in the source-drain electric field and produce dominating contributions to many kinetic coefficients. As a result the layer mobility, the Hall factor, the magnetoresistance, and the thermopower become anomalously large and dependent on the width of the STO sample and its bulk relaxation time.

$\phantom{}$
\vspace*{2ex} \par \noindent
{\em Acknowledgments.}

We are grateful to E. Baskin, A. V. Chaplik, M. I. Dyakonov, M. V. Entin, B. Jalan, A. Kamenev, C. Leighton, A. J. Millis, V. S. Pribiag, and S. Stemmer for helpful discussions. This work was supported primarily by the National Science Foundation through
the University of Minnesota MRSEC under Award No. DMR-1420013.

\bibliography{papers22}

\end{document}